# Laser induced surface nitriding of niobium: phase evolution and superconducting behaviour

J. Frechilla, A. Frechilla, G.F. de la Fuente, A. Larrea, L.A. Angurel, E. Martínez*

Instituto de Nanociencia y Materiales de Aragón, INMA, CSIC-Universidad de Zaragoza, María de Luna, 3, 50018 Zaragoza, Spain

## Abstract

Laser nitriding represents a versatile approach for tailoring the surface properties of metals. Up to now, its effect on the superconducting response of niobium nitrides remains largely unexplored. In this work, the nitriding process of niobium by laser irradiation under a controlled nitrogen atmosphere up to 2.50 bar, using a nanosecond pulsed laser with wavelength of 1064 nm has been investigated. By independently tuning the nitrogen pressure, the two-dimensional accumulated fluence ($F_{2D}$) and the laser irradiance, a laser-processing map for the formation of either a combination of β-$Nb_2N$ (hexagonal) and γ-$Nb_4N_{3\pm x}$ (tetragonal) phases or only the β-phase has been established. Systematic analysis by X-ray diffraction, scanning electron microscopy and electron backscatter diffraction revealed that the nitrogen-rich γ-phase forms in the near-surface layer through melting when $F_{2D}$ exceeds a certain value (> 50 kJ/cm$^2$ at 2.50 bar). A β-layer is observed underneath, and further inside, there is a band of embedded β-grains in the Nb matrix. Their size gradually decreases with increasing distance to surface, suggesting thermal gradients and a diffusion formation mechanism. When the γ-phase becomes predominant, a significant increase in the superconducting critical temperature is observed, up to $T_c \approx 15$ K, and magnetic irreversibility. For low $F_{2D}$ values (≈ 7.5 kJ/cm$^2$ at 1.50 - 2.50 bar), the formation of a uniform nitride layer composed by sub-micron-sized β-$Nb_2N$ grains results in a ca. fourfold enhancement in surface microhardness. These findings provide fundamental insights into laser-induced nitriding of niobium to engineer mechanically robust and superconducting Nb–N layers.

**Keywords:** Laser nitriding, Niobium nitride, γ-$Nb_4N_{3\pm x}$, β-$Nb_2N$, Microhardness, Superconductivity

*Corresponding author. E-mail: elena.martinez@csic.es

# 1. Introduction

Metal nitrides are of great interest due to their excellent hardness and thermochemical stability. For this reason, surface nitriding of pure metals and alloys has been widely used to increase their wear and corrosion resistance. Laser nitriding offers a number of advantages vs. traditional and plasma nitriding processes. These include surface-localized treatments which avoid the use of large volume furnaces or chambers, fast processing times (seconds or minutes vs. hours or days), high dimensional stability and energy efficiency. To our knowledge, it was first reported in the scientific literature by Katayama et al. for laser nitriding of Ti and its alloys [1], although patents were filed almost simultaneously in time for Al [2], and much later for steel [3]. Nitride processes have been applied to many metals, such as steel [4], Ti [5–8], Ti alloys [9, 10], CoCrFeMnNi [11] or Al [7, 8], for their use in a wide range of applications, such as engines [11], space industry [5] or biomedicine [6, 10].

Laser processing, which constitutes nowadays a widespread method for material surface modification, has also been used with this aim [4, 6–8, 10, 12, 13]. Laser nitriding is based on the irradiation with laser pulses of the material surface in a controlled nitrogen atmosphere, thus promoting metal nitride formation through diffusion or localized melting [14, 15]. The interaction between the heated surface and the laser-induced atmospheric plasma constitutes the primary mechanism driving effective nitridation [7]. This technique offers relevant features, including spatial precision, selectivity, and a relatively simple experimental setup, making it suitable for industrial applications [8]. For each material, the control of the thickness, phases and microstructural properties of the generated nitride coating layer implies an exhaustive analysis of the joint effect of the laser processing conditions, such as pulse duration ($\tau_{pulse}$), pulse energy ($E_{pulse}$), average pulse fluence ($F_{pulse}$), pulse repetition frequency ($f_{rep}$), irradiance ($I$) and overlapping between consecutive pulses. For example, Carpene et al. [7] reported excellent efficiency of laser nitriding of Ti from femtosecond (fs) to the nanosecond (ns) regimes, whereas for Fe, Al and steel, it improved in the ns range. The laser wavelength, $\lambda$, was a secondary parameter for all these metals. On the other hand, the nitrogen pressure, $P_{N2}$, is usually a key condition that affects this process [4, 8].

Besides the above mentioned characteristics, some metal nitrides exhibit very interesting functional properties. For example, some Nb and Ti nitrides are superconductors; ScAlN is ferroelectric; and Al, Ga and In nitrides are semiconductors, covering a wide-bandgap range [16]. With regard to the Nb–N system, which is the focus of the present work, it exhibits a wide variety of stable and metastable phases, including β-$Nb_2N$ (hexagonal), γ-$Nb_4N_{3\pm x}$ (tetragonal), ε-NbN (hexagonal), δ-NbN (fcc), as well as nitrogen-rich phases such as $Nb_5N_6$ and $Nb_4N_5$. These phases exhibit superconducting critical temperatures ($T_c$) higher and lower than that of pure Nb ($T_c \approx$ 9.25 K [17]) [18]. Reported values include $T_c <$ 2 K for β-$Nb_2N$ [19], $T_c \approx$ 11 K for ε-NbN [20], $T_c \approx$ 12–15 K for tetragonal γ-$Nb_4N_{3\pm x}$ [21, 22], while δ-NbN exhibits the highest critical temperature, with $T_c$ in the range of 15–16.5 K [21,23] depending on stoichiometry, reaching up to 17.3 K [24]. In addition, nitriding could also improve the mechanical properties of niobium [25].

Based on these properties, niobium nitrides have found applications across a range of technological fields, from microelectronics to advanced materials. Notable applications include superconducting microwave devices and radio-frequency (RF) cavities [26, 27], single-photon and X-ray superconducting detectors [28, 29], high-energy-density supercapacitors [30], superconducting quantum interference devices (SQUIDs) [31, 32], and emerging quantum technologies [33]. In addition, the high hardness and wear resistance of these phases make niobium nitrides suitable for durable protective coatings [34].

The synthesis of these phases has been achieved using various techniques. The most common approach involves direct reaction of niobium with nitrogen at high temperatures in furnaces or through diffusion treatments [35–41]. Alternative physical and chemical methods have also been explored, including pulsed laser deposition (PLD) [42–44], magnetron sputtering [45], chemical vapour deposition (CVD) [46], atomic layer deposition (ALD) [47] or plasma-assisted nitriding [48]. Laser nitriding of Nb has already been reported by other groups using ns [12, 13] and fs [13] pulsed lasers. However, the superconducting properties of the resultant materials were not reported.

The objective of the present work is to investigate the laser nitriding of niobium by irradiating Nb sheets with a ns pulsed-laser with emission wavelength $\lambda$ = 1064 nm in a chamber under controlled nitrogen pressure. Phase and microstructural analysis of the generated nitrides were combined with their superconducting behaviour for a wide range of laser parameters and pressure values. This enables to identify the range of laser conditions that achieve optimal mechanical performance and the foreseen $T_c$ increase.

## 2. Experimental

### 2.1. Sample preparation and laser nitriding method

Commercial 1mm-thick niobium sheets (purity 99.9%, Goodfellow NB000380) were cut into 10 mm × 10 mm pieces using a pulsed fs green laser (Light Conversion, Vilnius, Lithuania, Carbide CB3-40 W, $\lambda$ = 515 nm, $\tau_p$ = 249 fs) operating at a pulse repetition frequency of $f_{rep}$ = 200 kHz and a scanning speed of 30 mm/s. After cutting, the samples were ultrasound cleaned in isopropanol for 15 minutes. The sample surfaces were then mechanically polished using progressively finer sandpapers (from P180 to P2500) to achieve a smooth and homogeneous surface suitable for subsequent laser treatments. Following polishing, the samples were again ultrasound cleaned in isopropanol for 15 minutes to remove any residual contaminants from the sanding process.

For laser nitriding, the samples were placed in a sealed chamber, which was vacuumed to $\approx 8 \cdot 10^{-3}$ mbar before introducing nitrogen gas (5 N purity, Nippon Gases) to the set pressure, between 0.15 and 2.50 bar. The niobium samples were then irradiated with an Ytterbium pulsed fibre laser (PEDB-400B, Perfect Laser Co., Ltd., Wuhan, China), emitting at a central wavelength of $\lambda$ = 1064 nm and a maximum power of 70 W, by using pulse widths ranging between 20 and 200 ns and pulse repetition frequencies between 175 and 1000 kHz. At a working distance of 163 mm, the laser beam exhibited a Gaussian spatial energy profile with a circular spot diameter $2r_b$ = 65 µm for the 1/e² decay distance. The beam was moved at speed $v_B$ in a single direction over the sample surface. To scan the entire surface, either the laser beam scanned a meander-type path with a hatching parameter $\delta$ between consecutive lines, or the sample was moved at a speed $v_S$ in the direction perpendicular to the beam scan direction (Laser Line Scanning, LLS, method [49, 50]).

*2.2. Phase and microstructure characterization*

Phase formation of niobium nitrides was analysed by X-ray diffraction (XRD) using a PANalytical Empyrean system in Bragg–Brentano geometry, equipped with a Cu Kα source (wavelength of 1.5418 Å) and a PIXcel linear detector. The 2θ range of 30–80° was analysed to capture the complete signal from the nitride layer.

The microstructure characteristics of the generated nitride layers (morphology, phases and crystallography) were observed using a MERLIN field-emission scanning electron microscope (FESEM, Carl Zeiss, Jena, Germany). Electron Back Scattering Diffraction (EBSD) experiments were performed with the same microscope using an AztecHKL system from Oxford Instruments (Abingdon, UK) to analyse the surface crystallography. The FESEM was operated between 5 and 20 kV, using angle selective backscatter (AsB) and forescatter electron (FSE) detectors.

For SEM (AsB and EBSD) analysis, the samples were embedded in epoxy resin and polished to analyse their cross-sections perpendicularly to the beam scan direction. The polishing was performed in several steps using sandpapers, diamond and colloidal silica, with a final cleaning step to remove the silica. The samples were ultimately coated with approximately 5 nm of carbon because the used epoxy resin is not electrically conductive.

*2.3. Hardness tests*

Microhardness measurements were performed using a Vickers tester (MXT50, Matsuzawa Seiki Co., Tokyo, Japan). Loads ranging from 25 g to 500 g were applied, with a dwell time of 15 s, to assess hardness at different depths and obtain a depth-dependent hardness profile. Each measurement was repeated three times per load to minimize variability, and the mean and standard deviation were calculated.

*2.4. Characterization of superconductivity properties*

Superconductivity property characterization was carried out in a SQUID magnetometer (MPMS3, Quantum Design, San Diego, CA, USA). The critical temperatures were obtained from AC magnetic susceptibility, $\chi_{ac}$ ($T$), including in-phase ($\chi'$) and out-of-phase ($\chi''$) components, measured in zero DC magnetic field with an AC driving field of $\mu_0 h_0$ = 0.1 mT at $f$ = 10 Hz. $T_c$ was determined as the onset of diamagnetism, i.e. at the temperature where the modulus of $\chi'$ exceeds the noise level. In addition, the magnetic irreversibility and flux-pinning behaviour was analysed by zero-field-cooling (ZFC) and field-cooling (FC) (under DC fields of 2 mT) magnetization curves and by isothermal magnetic hysteresis loops, $M(\mu_0 H)$, at different temperatures above the critical temperature of pristine Nb (9.25 K). For all the measurements, the magnetic field was applied perpendicular to the sample surface.

## 3. Results

*3.1. Effect of laser processing parameters in the nitriding process*

Several sets of Nb samples were processed using different conditions of pressure, average pulse fluence, irradiance and overlap between consecutive laser pulses. XRD patterns were initially obtained from the surfaces of all these samples. This enabled a systematic analysis of the influence of the laser parameters on the nitriding process, particularly regarding the appearance of the different crystalline phases in the Nb-N system. Table 1 collects some selected representative samples from this study. The accumulated 2D fluence is defined here as, $F_{2D} = F_{pulse} \cdot [\pi \cdot r_b^2/(\delta_p \cdot \delta_l)]$, which depends on the distance between consecutive pulses in a line ($\delta_p = v_B/f_{rep}$) and between lines, $\delta_l$ [51].

Figure 1 shows XRD patterns obtained on several of these samples. Figure 1A corresponds to samples irradiated under different nitrogen pressures at low accumulated fluence values of $F_{2D} \approx 7.5$ kJ/cm$^2$. This condition just exceeds the nitriding threshold of niobium at a pressure of 0.15 bar (sample S1), where the onset of the hexagonal β-Nb$_2$N phase is detected. The corresponding peaks' intensity increases with pressure. This experiment also reflects a change in the orientation of the β-Nb$_2$N phase upon increasing the nitrogen pressure. Specifically, for $P_{N2} \geq 1.50$ bar, the intensity of the peak corresponding to the (002) reflection decreases, while that of the (100) increases (note that both peaks exhibit similar intensity values for non-textured samples [52]). The observed crystallographic texture would be associated to the thermal gradients induced on the surface and through the thickness of the samples by the nanosecond laser pulses [53–55].

Figure 1B shows the evolution observed in the XRD patterns when $F_{pulse}$ and $F_{2D}$ values were progressively increased, at $P_{N2}$ = 2.50 bar (the maximum pressure allowed by our experimental set up). Note that the peak associated to the tetragonal γ-Nb$_4$N$_{3\pm x}$ phase, which just emerged at 7.5 kJ/cm$^2$, becomes clearly visible for all these conditions. For $F_{2D}$ = 56 and 113 kJ/cm$^2$ (samples S5 and S6, respectively), the intensity of both nitride phases reflections further increases, as compared to the samples shown in Figure 1A. At $F_{2D} \approx 528$ kJ/cm$^2$ (sample S7), the (200) and (004) reflections of γ-Nb$_4$N$_{3\pm x}$ become fully distinguishable and, at the same time, the intensity of the β-Nb$_2$N peaks starts decreasing. For $F_{2D} \approx 1056$ kJ/cm$^2$ (sample S9), the tetragonal nitride phase remains dominant. It should be reminded here that the γ-Nb$_4$N$_{3\pm x}$ phase is a tetragonal distortion of the fcc unit cell of the δ-NbN$_{1-x}$ compound and exhibits compositional variations, usually between 40-45 at% N (the latter value corresponding to a ratio N/Nb ≈ 0.82) [39–41]. The splitting of the fcc diffraction lines, indicative of formation of this tetragonal phase, has been reported in the literature for bulk niobium samples after a nitriding process using furnaces [38, 39]. In those studies the presence of a δ+γ band was observed, which would form as a consequence of the partial transformation of δ-NbN$_{1-x}$ → γ-Nb$_4$N$_{3\pm x}$ upon cooling for N/Nb < 0.84 [39, 41].

**Table 1**. Selected laser processing conditions employed for nitriding of niobium sheet samples using $\tau_{pulse}$ = 200 ns.

| Name | $F_{pulse}$ (J/cm$^2$) | $f_{rep}$ (kHz) | $v_B$ (mm/s) | $\delta_L$ (µm) | $P_{N2}$ (bar) | $I$ (GW/cm$^2$) | $F_{2D}$ (kJ/cm$^2$) |
|---|---|---|---|---|---|---|---|
| S1  | 2.41  | 700 | 250 | 3    | 0.15 | 0.0121 | 7.5  |
| S2  | 2.41  | 700 | 250 | 3    | 0.50 | 0.0121 | 7.5  |
| S3  | 2.41  | 700 | 250 | 3    | 1.50 | 0.0121 | 7.5  |
| S4  | 2.41  | 700 | 250 | 3    | 2.50 | 0.0121 | 7.5  |
| S5  | 6.03  | 350 | 125 | 1    | 2.50 | 0.0301 | 56   |
| S6  | 12.05 | 175 | 60  | 1    | 2.50 | 0.0603 | 113  |
| S7  | 12.05 | 175 | 30  | 0.45 | 2.50 | 0.0603 | 528  |
| S8  | 12.05 | 175 | 30  | 0.35 | 2.50 | 0.0603 | 660  |
| S9  | 12.05 | 175 | 30  | 0.22 | 2.50 | 0.0603 | 1056 |
| S10 | 12.05 | 175 | 15  | 0.26 | 2.50 | 0.0603 | 1761 |
| S11 | 12.05 | 175 | 15  | 0.18 | 2.50 | 0.0603 | 2640 |

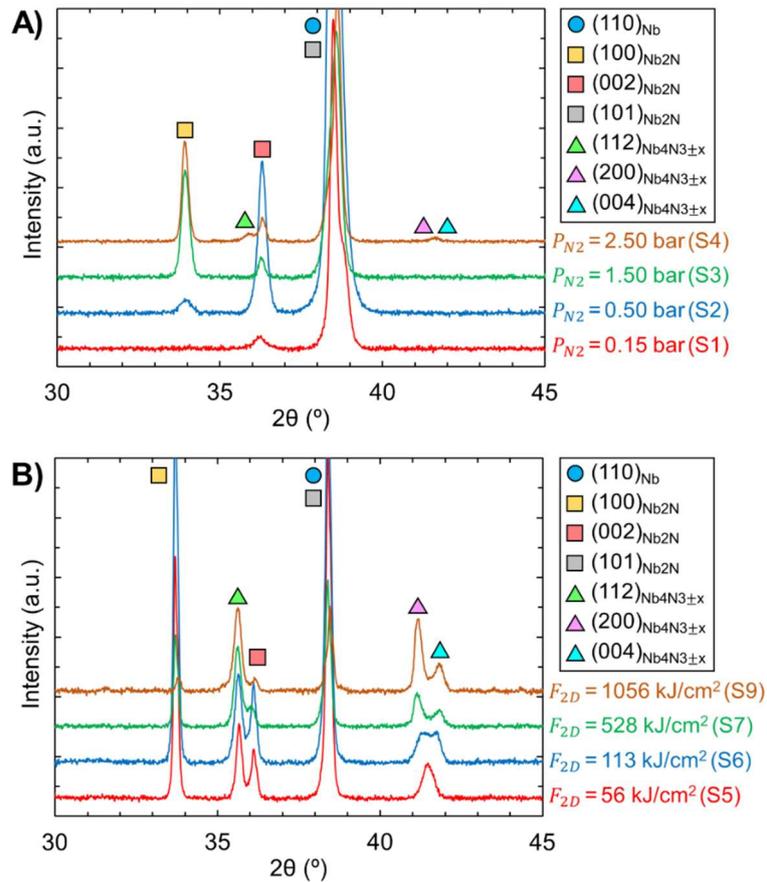

**Figure 1**. XRD patterns of selected laser-processed samples (see table 1) showing the evolution of the niobium nitride phases by modifying A) the nitrogen pressure for $F_{2D}$ = 7.5 kJ/cm$^2$; and B) the 2D accumulated fluence for $P_{N2}$ = 2.50 bar. The peaks corresponding to Nb (circles), β-Nb$_2$N (rectangles) and γ-Nb$_4$N$_{3\pm x}$ (triangles) are indicated. Note that the (101)$_{Nb2N}$ is almost coincident with the (110)$_{Nb}$. The complete measured XRD patterns (angular range between 30° and 80°) are provided in the Supplementary Information (section S1).

To gain further insights into the mechanisms governing phase formation, we have calculated the fraction of the peak intensities $(112)_{Nb4N3\pm x}$ and $(002)_{Nb2N}$ for different laser processing conditions. The results are summarized in Figure 2. Figure 2A shows the evolution upon increasing the $F_{2D}$ value for a fixed value of irradiance ($I \approx 0.06$ GW/cm$^2$, with $F_{pulse} \approx 12.05$ J/cm$^2$, $\tau_{pulse}$ = 200 ns and nitrogen pressure of 2.50 bar). The increase of pulse overlapping clearly promotes nitrogen diffusion and the formation of γ-Nb$_4$N$_{3\pm x}$, which reaches a maximum at $F_{2D} \approx 1760$ kJ/cm$^2$ (sample S10). When $F_{2D}$ is increased beyond this point, a small reduction in this ratio is observed. Figure 2B shows the effect of changing the laser pulse irradiance. This was done by varying the pulse width from $\tau_{pulse}$ = 20 ns to 200 ns while fixing the pulse fluence and overlap conditions ($F_{pulse} \approx 4.82$ J/cm$^2$ or 6.03 J/cm$^2$, $\delta_p \approx 0.35$ μm, $\delta_L$ = 3 μm). In these experiments the pressure was set at 1.50 bar. Two distinct regions are identified. For $I \leq 0.1$ GW/cm$^2$ there is a significant presence of γ-Nb$_4$N$_{3\pm x}$, and it sharply decreases at higher values of irradiance. This reduction would be likely due to the onset of ablation, which removes progressively the nitrided layer during processing, as observed in other metals [56].

To further analyse the range of laser processing conditions that result in the formation of different niobium nitride phases for the processed samples, Figure 2C shows a two-dimensional map of $I$ versus $F_{2D}$. Each point corresponds to a particular experiment performed at 1.50 bar (circles) or 2.50 bar (rectangles). Different symbols are used depending on the phases observed in XRD experiments: no nitrides (open symbols), only the β-Nb$_2$N phase (semi-full) or β-Nb$_2$N + γ-Nb$_4$N$_{3\pm x}$ phases (full symbols). The background regions shaded in colours are a guide to the eye for better visualization of each case. The area shadowed in darker blue corresponds to the conditions where the double peaks characteristic of the tetragonal phase were well visible in the XRD pattern. This representation highlights the predominant influence of the $F_{2D}$ variable over irradiance to obtain the foreseen nitride phases with higher $T_c$ values than Nb.

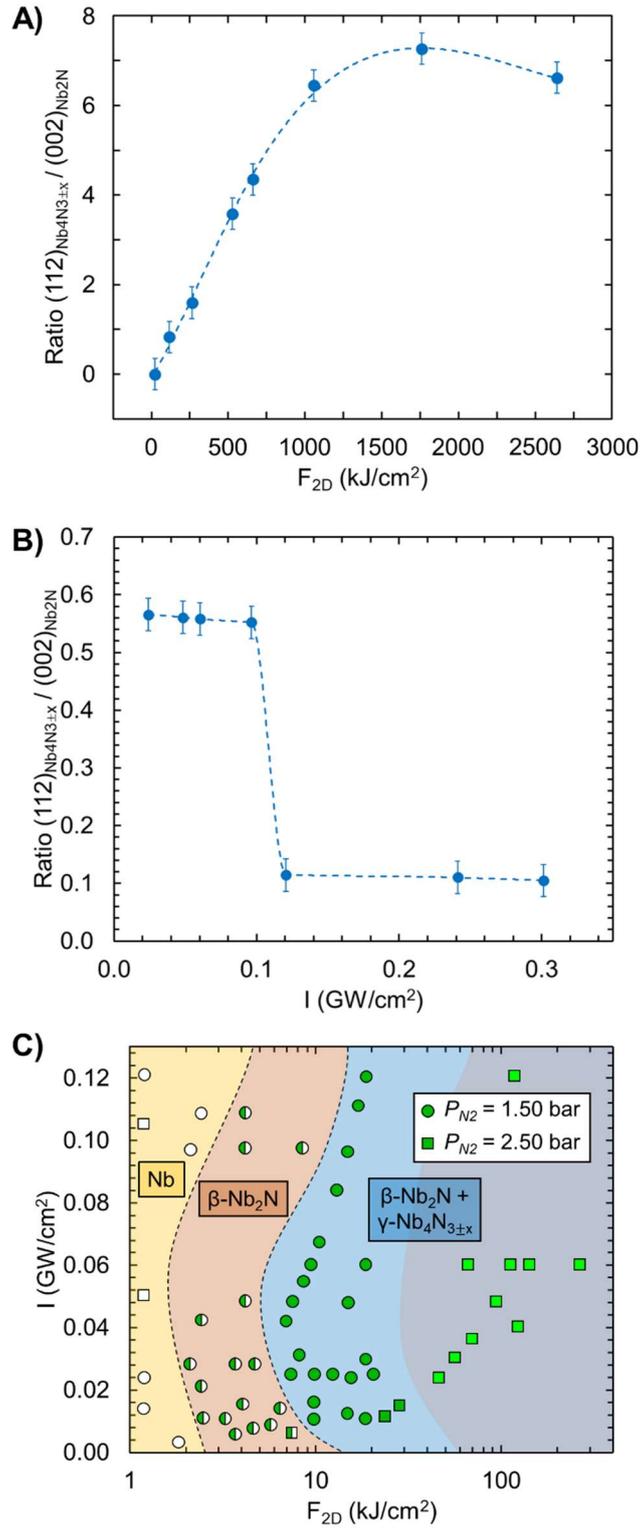

**Figure 2.** A) $(112)_{Nb4N3\pm x}/(002)_{Nb2N}$ intensity ratio as a function of $F_{2D}$ at constant irradiance ($F_{pulse} \approx 12.05$ J/cm$^2$, $\tau_{pulse}$ = 200 ns, $I \approx 0.06$ GW/cm$^2$). Dashed lines are guides for the eye. B) $(112)_{Nb4N3\pm x}/(002)_{Nb2N}$ values as a function of the irradiance at constant $F_{pulse}$ = 4.82 J/cm$^2$ or 6.03 J/cm$^2$, $\delta_p \approx 0.35$ μm and $\delta_L$ = 3 μm. C) 2D map of irradiance versus two-dimensional fluence. Each point represents a nitriding experiment (open symbols: no nitrides observed in XRD, semi-full: just β-Nb$_2$N; full: β-Nb$_2$N+ γ-Nb$_4$N$_{3+x}$). The background colours are a guide to the eye to better visualize these regions. Circles and rectangles correspond to pressures of 1.50 or 2.50 bar, respectively.

*3.2. Microstructural properties of the laser-nitrided layers*

The spatial distribution of the different nitride phases generated by laser has been analysed by FESEM and EBSD experiments. With this aim, some samples exhibiting in XRD patterns only the β-$Nb_2N$ phase (S3) or both β-$Nb_2N$ and γ-$Nb_4N_{3\pm x}$ phases (S5, S6, S7 and S9, ordered by increasing intensity of the tetragonal phase) were selected for this study.

Figure 3 shows FESEM micrographs corresponding to the polished cross-sections. The thickness of the nitride layer tends to increase with the accumulated fluence. Thus, sample S3 exhibits about 4 μm–thick band of β-$Nb_2N$ grains embedded in the niobium (lighter grey colour in the AsB images). On the surface, these grains form a quasi-continuous layer with ≈0.4-0.5 μm of thickness. Upon increasing the $F_{2D}$ value, the affected area penetrates deeper into the bulk and the nitride grains increase in size. The thickness of the outermost nitride-layer reaches values of ≈2.0±0.5, 2.5±0.5, 6.0±2.0 and 4.5±2.0 μm (for S5, S6, S7 and S9, respectively). Near the surface of samples S7 and S9, two sub-regions within the continuous layer are distinguished, exhibiting contrast differences due to compositional and/or crystallographic variations. This suggests the local coexistence of β-$Nb_2N$ and γ-$Nb_4N_{3\pm x}$ domains, which will be analysed with more detail by the EBSD technique.

The underlying band of niobium nitride grains dispersed within the Nb metallic matrix also appeared in bulk niobium nitrided using induction heating and conventional furnace processing [38, 40, 41]. Their formation originates from the precipitation of β-$Nb_2N$ inside the α-Nb(N) solid solution during cooling, caused by the lower solubility of nitrogen in niobium as the temperature decreases. The fact that their size diminishes along the sample's thickness in our laser-nitride samples is a consequence of the thermal gradient inherent to this technique with enhanced nitrogen diffusion and greater solute trapping near the surface. The thickness of this region also increases with the 2D fluence. For example, the micrographs of sample S7 show that the grain size decreases from a few micrometres (near the outer nitride layer) down to the few nanometres (at about 30 μm distance from the surface).

Some samples show features of laser melting, especially when $F_{2D}$ exceeds ≈50 kJ/$cm^2$ (at 2.50 bar). Microcracks are also observed, attributed to residual stress accumulation caused by rapid cooling and differential contraction between the nitride layer and the niobium core. Conventional nitriding methods reported in the literature for Nb are usually performed at temperatures below the melting temperature of Nb [39–41]. However, this normally requires considerably longer processing times and/or pressure (for example, 330 h at 1200 °C, < 100 mbar [39]; 2h at 1777 °C, 30.5 bar [40]). The pressure limits of our set-up likely explains the need to exceed locally the Nb melting temperature to generate the foreseen high-$T_c$ nitrides, given the short reaction times that characterise laser processing.

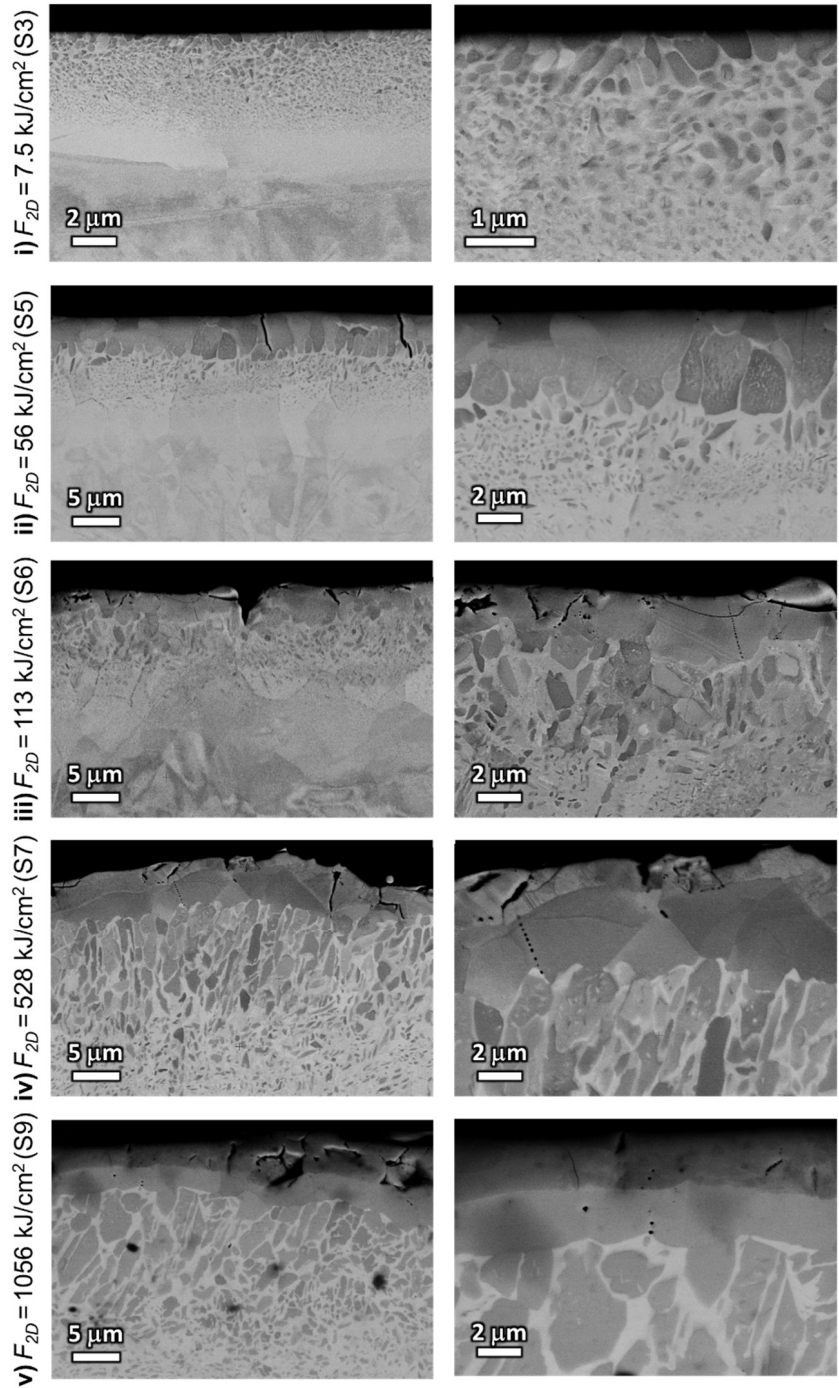

**Figure 3.** FESEM (AsB) images of the polished cross-sections of laser-nitrided niobium samples processed at different conditions (see table 1), taken at two different magnifications. Note that images of sample S3 are presented at a higher magnification to allow clearer visualization of the nitride layer.

A detailed analysis of the distribution of the laser-generated nitrides was performed by EBSD experiments on polished cross-sections. Figures 4 and 5 display the maps corresponding to Phase and Inverse Pole Figure (IPF) for samples S6, S7 and S9. These samples were selected because both β and γ nitride phases were detected in XRD experiments. It is important to recall that IPF maps differentiate grains by visualizing crystallographic orientations relative to a sample direction. Thus, these maps enable direct correlation between phase distribution, crystallographic orientation and local deformation within the nitride layers and the Nb matrix. In particular, IPF-Y map was represented in Figures 4 and 5, as an example, with Y-direction parallel to the vertical orientation in these images. Black points in these maps represent zero solution data points, which are generally located around grain boundaries. Further technical details of the EBSD analysis, including map acquisition parameters, indexing conditions as well as the complete EBSD maps, are provided in the Supplementary Information (section S2).

Figure 4 displays the collected maps for sample S6, which was processed with $F_{2D}$ = 113 kJ/cm$^2$. The Phase-map shows that the nitride layer in this sample is composed predominantly of β-Nb$_2$N phase (in red colour). The β-grains form an almost continuous layer near the surface and are also embedded within the Nb (yellow) matrix. On the surface, a submicron-thick, discontinuous γ-Nb$_4$N$_{3\pm x}$ (green) layer indicates local stabilization of the nitrogen-rich tetragonal phase under conditions of enhanced nitrogen availability.

By comparing the IPF and Phase maps, we observe that the upper β-layer is formed by grains of size typically in the range 0.6 – 3 µm (measured perpendicularly to the surface). Their size gradually decreases with the distance to surface, as mentioned above. The mean aspect-ratio of β-grains is about 2. These maps also reveal that the inward diffusion of nitrogen takes place preferentially through grain boundaries, where interstitial nitrogen diffusion is enhanced. As a consequence, β-precipitates extend through the boundaries and are also formed inside the Nb grains.

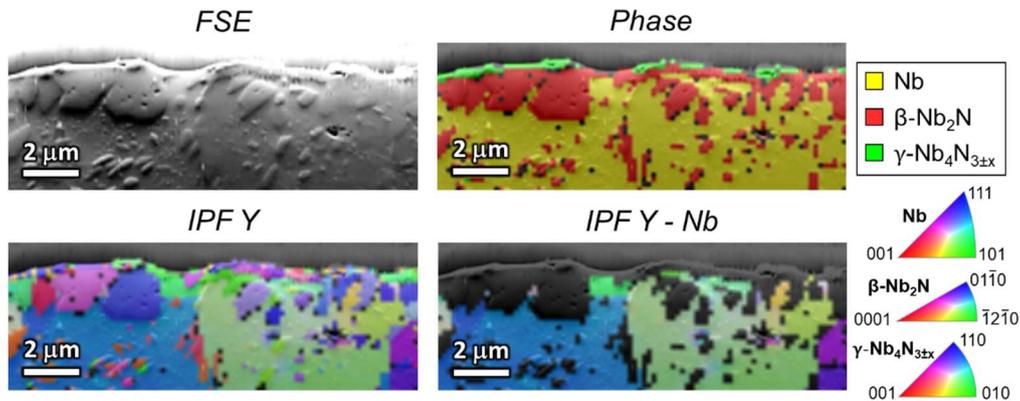

**Figure 4.** EBSD maps of polished cross-section of the laser nitrided sample S6 corresponding to: *Forescatter Electron* images (FSE), *Phase* and *Inverse Pole Figure* (IPF-Y), as indicated. The IPF-Y map of all phases (left) and just Nb (right) is displayed for better understanding.

The corresponding maps for samples S7 and S9, processed with higher 2D-fluences, are shown in Figure 5. Compared to S6, it is noted that the γ-phase of sample S7 becomes more prominent and extends deeper into the nitrided region, coexisting with β-domains. Sample S9 exhibits a thicker γ-layer (≈2–3 µm), with a relative fraction of the tetragonal phase larger than for the S7 sample, consistent with the XRD results shown in Figure 1B. IPF maps also reveal differences in the generated γ-Nb$_4$N$_{3\pm x}$ layer between both samples. The γ-grains in sample S7 display irregular morphologies, diffuse interfaces and a broad orientation dispersion, consistent with rapid solidification and competitive nucleation. On the contrary, in sample S9 (processed with a higher $F_{2D}$ value), the tetragonal γ-grains appear more sharply defined and better delineated, indicating improved crystallographic coherence. This suggests that the higher thermal input promotes partial recrystallization and stabilization of the tetragonal phase, enabling grain growth, especially in the direction parallel to the surface, and reducing interfacial disorder.

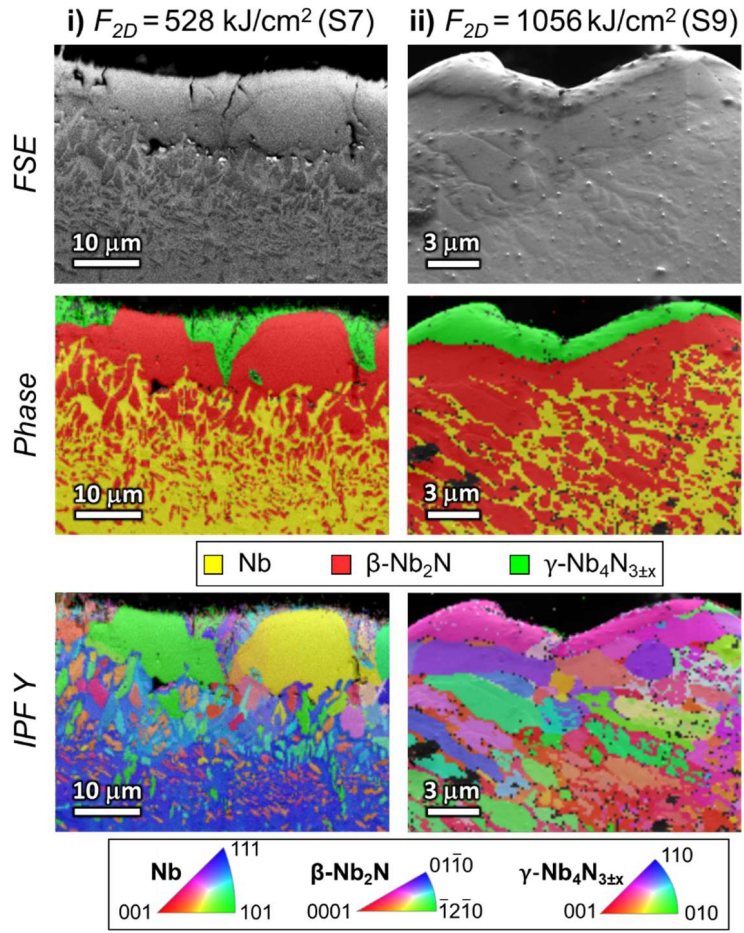

**Figure 5.** EBSD maps –FSE, Phase and IPF-Y, from top to bottom– of polished cross-sections of samples i) S7 and ii) S9.

EBSD analysis confirms the absence of the cubic δ-$NbN_{1-x}$ phase in all these samples. As already mentioned in section 3.1, this is indicative of nitrogen deficiency at the surface for the temperature/pressure conditions used in the present work [41]. The formation of the δ-phase at temperatures near or above the melting temperature of Nb, as in this case, would require higher nitrogen pressure during laser processing, likely of the order of 10 bar. This is consistent with published work of niobium nitridation using conventional furnaces or combustion synthesis methods [38, 41].

*3.3. Mechanical properties of the nitrided layers*

Figure 6 shows the Vickers microhardness experiments performed in different samples to analyse the changes in the mechanical properties produced by the generated nitride phases. The curves display the variation of hardness with indentation depth, thereby probing sequentially from the outermost surface to the deeper regions, with penetration depths of approximately 3–25 µm. This study was mainly focused on the samples irradiated with low $F_{2D}$ values, where the β-$Nb_2N$ phase was primarily detected (see Figure 1A). Literature reports similar hardness values for β-$Nb_2N$ and γ-$Nb_4N_{3\pm x}$, (in the range 12–26 GPa, depending on the processing and measurement method [57–59]), suggesting that the presence of γ-$Nb_4N_{3\pm x}$ would not provide additional mechanical enhancement. Furthermore, the lower processing temperatures of S1–S4 samples produce fewer surface defects and cracks than in samples with mixed phases, as noted above, resulting in more reliable indentation measurements.

As observed in Figure 6, the surface microhardness notably increases with the formed nitride thickness, reaching an approximately fourfold enhancement compared to the pristine material. The relatively low hardness values measured in this work (≈ 3.5 GPa for sample S4), compared to literature reports, can be attributed to the nitride grain size heterogeneity across the layer thickness and to the effect of the Nb substrate under certain indentation conditions. In this regard, the cross-section of sample S3 shown in Figure 3i reveals a nitrided layer thickness of about ≈ 4 µm, comparable to the estimated penetration depth at the lowest applied load for sample S3 (≈ 3.9 µm). This suggests that even low-load indentations probe a significant portion of the nitrided layer.

As the penetration depth increases, the measured hardness progressively reflects the contribution of the underlying niobium substrate, leading to a reduction in the apparent hardening effect. Therefore, the formation of the β-$Nb_2N$ phase significantly improves the surface hardness, making it suitable for durable protective coating applications.

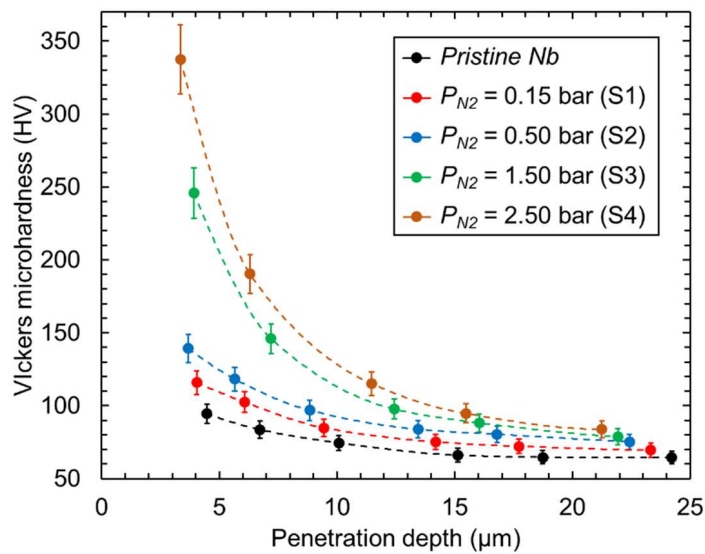

**Figure 6.** Vickers microhardness as a function of applied load, measured on the surface of the samples processed with $F_{2D}$ = 7.5 kJ/cm$^2$ under different nitrogen pressures (see table 1).

### 3.4. Superconductor properties of the niobium nitrided samples

Figure 7 summarizes the measured superconductor properties of some laser-nitrided samples. These samples were selected because of the presence of the γ-Nb$_4$N$_{3\pm x}$ phase by XRD. A pristine Nb plate was also measured for comparison. Figure 7A displays the in-phase component of the AC susceptibility as a function of temperature for different samples. These measurements were used to estimate the critical temperature of bulk and laser-nitride layers. A sharp decrease of the diamagnetic signal is observed in all samples at 9.25 K, coincident with the $T_c$ of pristine Nb ($T_{c,Nb}$). Except for sample S6, all analysed samples exhibit an increase of $T_c$ to approximately 14.75 ± 0.25 K, consistent with the expected values for this phase. Moreover, the diamagnetic signal above $T_{c,Nb}$ increases with the relative amount of γ-phase, with sample S10 ($F_{2D}$ = 1761 kJ/cm$^2$) exhibiting the highest fraction of this phase (see Figure 2A).

Contrary to what was expected, no superconducting signal above $T_{c,Nb}$ was detected in sample S6, despite presenting the tetragonal γ-phase in XRD and EBSD analysis. This could suggest that either a minimum degree of crystallographic ordering, nitride thickness layer and/or continuity, or a specific composition within the γ-Nb$_4$N$_{3\pm x}$ phase is necessary for the emergence of superconductivity in the nitride layer at temperatures higher than $T_{c,Nb}$. It is also noted that in our samples, this onset was observed when the tetragonal phase exhibits a complete splitting of the mentioned XRD peaks (note that the peak splitting is just beginning in sample S6, as seen in Figure 1B). The continuous and smooth decrease of the diamagnetic signal in the temperature range between $T_{c,Nb}$ and $T_c$ would indicate some intergranular or grain boundary effects in the nitride layer, which could be caused by a distribution of $T_c$ values or/and connectivity properties between grains (or group of grains) in addition to variations in stoichiometry [60, 61].

To analyse the magnetic irreversibility behaviour of the nitrided layers, magnetization measurements were performed on different samples. In particular, Figures 7B and 7C show, respectively, the ZFC-FC $M(T)$ curves under a field of 2 mT and the isothermal $M(\mu_0 H)$ hysteretic loops at 10 K, above $T_{c,Nb}$ in order to avoid any contribution from the bulk substrate. Since the effective thickness $d$ of the superconducting nitride layers is difficult to estimate accurately and varies between samples, the quantity $M \cdot d$ (equivalent to $m/(2a)^2$) is plotted in the figure, where $m$ is the magnetic moment and $a$ ≈ 1.5 mm, the semi-width of the measured square plates. This representation allows direct comparison of the induced magnetic moment without introducing additional uncertainty associated with the layer thickness.

As seen in Figure 7B, for all these samples the nitrided layers remain superconducting under a field of 2 mT at $T > T_{c,Nb}$, although a reduction in $T_c$ is observed in some of them. Thus, $T_c$(2mT) decreases to 13.75 ± 0.25 K for samples S7 and S8, but remains approximately equal to $T_c$ at zero field for samples S10 and S11, which contain a larger amount of the tetragonal γ-phase. In all samples FC and ZFC curves merge at the onset of diamagnetism, indicative of flux pinning behaviour just below $T_c$. It is noteworthy that the shape of ZFC-$M(T)$ and $\chi'(T)$ curves are very similar, despite the different magnetic fields sensing the samples in both measurements (2 mT vs 0.1 mT, respectively).

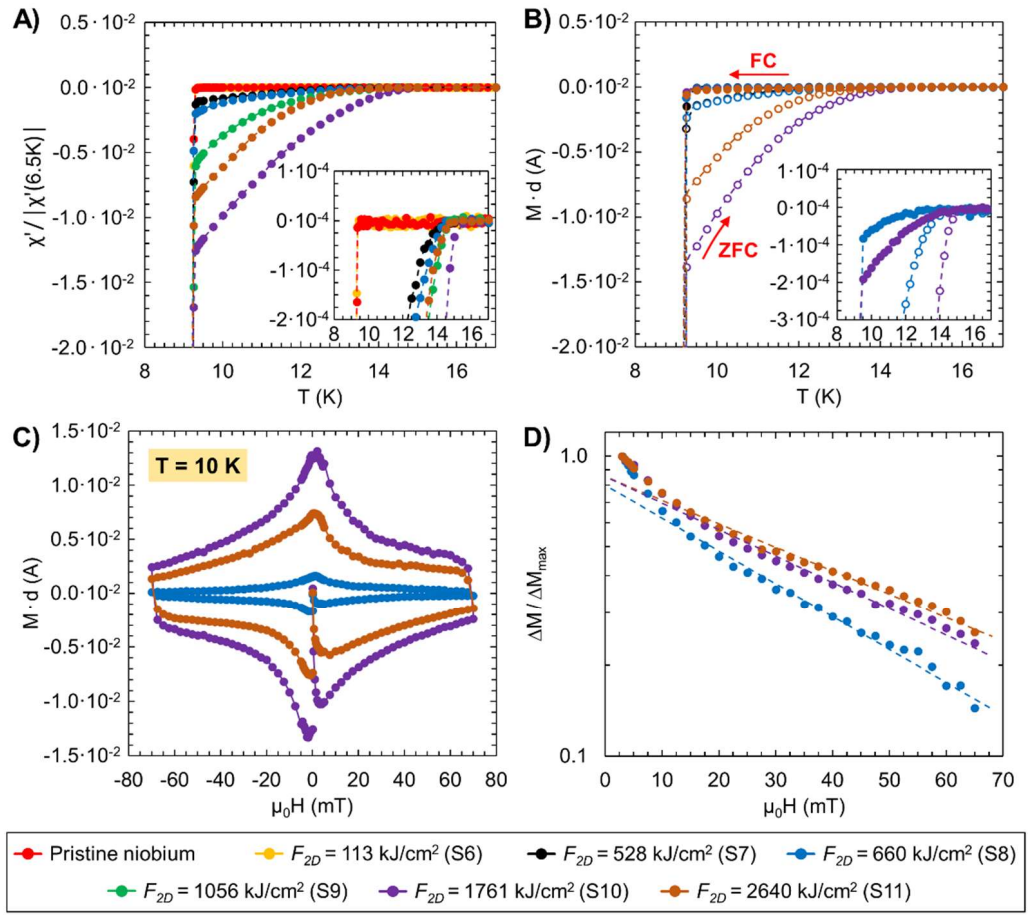

**Figure 7.** Superconducting properties measured on 2*a* x 2*a* processed samples at temperatures higher than $T_{c,Nb}$: A) AC susceptibility (in-phase component). B) ZFC–FC magnetization measurements (open-full symbols), plotted as $M \cdot d \equiv m/(2a)^2$, under an applied field of 2 mT (see text). C) Magnetization hysteresis loops, $M(\mu_0 H) \cdot d$, measured at 10 K and D) corresponding $\Delta M / \Delta M_{max}$ calculated from C). The discontinuous lines in D) correspond to exponential decay fitting between 10 and 70 mT, as explained in the text. The same colour-coding was used for the four figures.

The magnetization hysteresis loops $M(\mu_0 H) \cdot d$ at 10 K plotted in Figure 7C also display some differences among samples. Clearly evidence of irreversibility behaviour (vortex pinning) is observed in all these samples, even at fields as high as 70 mT. In consonance with the previous results, the highest magnetization signal is obtained for sample S10, followed by S11 and S8. Figure 7D shows the normalized magnetization loop width, $\Delta M/\Delta M_{max}$, as a function of the applied magnetic field, derived from the hysteresis cycles. Here, $\Delta M$ corresponds to the difference between the ascending and descending branches of the magnetization loops at each field, and $\Delta M_{max}$ their maximum value for each sample. It is observed that all curves exhibit exponential decays, $\Delta M \propto exp(-H/H^*)$, with characteristic decay fields $\mu_0 H^* \approx 40$ mT, 50 mT and 55 mT (± 1.5 mT) for samples M14, M16 and M17, respectively (in the magnetic field range between 10 and 70 mT at this temperature).

Figure 8 displays the magnetization hysteresis loops at different temperatures ($T > T_{c,Nb}$) for the sample that exhibits the best results, S10. The corresponding $\Delta M/\Delta M_{max}$ curves, normalized to the maximum value at 9.5 K, are also displayed. For all temperatures, an almost exponential decay of $\Delta M/\Delta M_{max}$ with applied field is observed, showing a decreasing trend with different slopes in distinct field regions. The decay rate is steeper with increasing temperature, as expected. A well-defined superconducting hysteresis loop is still clearly observed at 14 K, confirming the persistence of flux pinning even close to the critical temperature of the nitride layer.

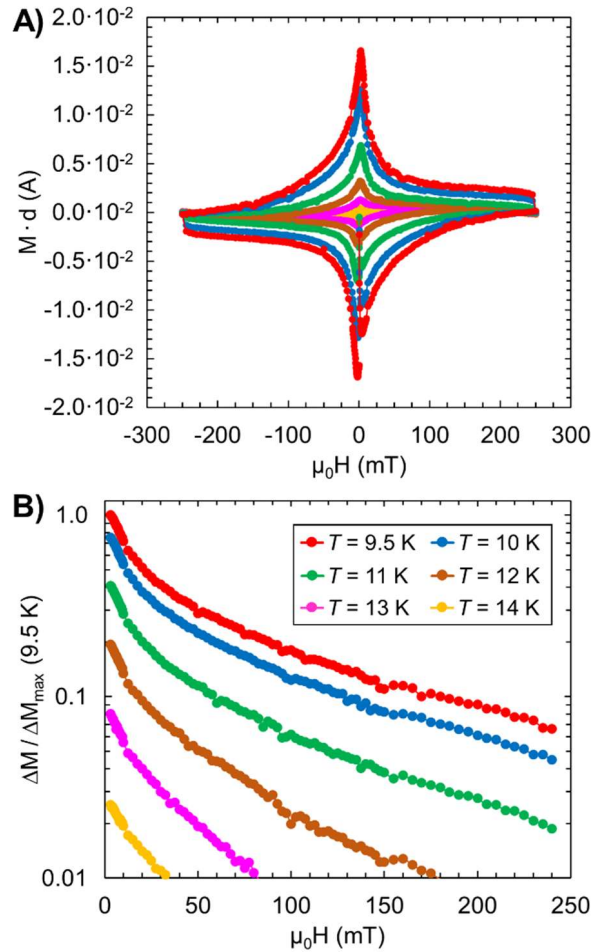

**Figure 8.** A) Magnetization hysteresis loops $M \cdot d$ ($\mu_0 H$) of sample S10 at different temperatures above the $T_{c,Nb}$. B) Corresponding $\Delta M/\Delta M_{max}$ (9.5K) curves.

## 4. Conclusions

The present work demonstrates that the irradiation of bulk niobium with a nanosecond pulsed laser at a wavelength of 1064 nm under controlled nitrogen atmosphere enables the formation of different niobium nitride phases on its surface. By adjusting the nitrogen pressure, the accumulated two-dimensional fluence ($F_{2D}$) and the irradiance, it is possible to selectively generate either a combination of β-$Nb_2N$ (hexagonal) and γ-$Nb_4N_{3\pm x}$ (tetragonal) phases or just the former. In particular, the increase in nitrogen pressure and $F_{2D}$ enhances the nitrogen diffusion into Nb, favouring the stabilization of nitrogen-richer phases. The phase with the highest nitrogen content consistently forms in the outermost region, where nitrogen availability is the highest. In all samples, the resulting cross-section comprises a continuous surface layer together with a diffusion-affected region containing isolated β-grains.

At low $F_{2D}$ values (typically lower than 10 kJ/cm$^2$ under a nitrogen pressure of 1.50 - 2.50 bar) the submicron-thin nitride layer composed by β-grains results in a significant enhancement of the surface microhardness. Thus, Vickers hardness up to four times higher than those of pristine niobium were reached, while maintaining the integrity of the surface. This confirms the potential of laser-generated β-$Nb_2N$ layers for durable protective coating applications.

The γ-phase forms at higher fluences, $F_{2D}$ > 50 kJ/cm$^2$ under 2.50 bar. These samples show clear microstructural features of Nb melting near the surface, followed by rapid solidification and nitrogen diffusion during the process. Samples processed with $F_{2D}$ > 500 kJ/cm$^2$ show an increase in the normal/superconducting transition temperature up to approximately 15 K, and exhibit magnetic irreversibility just below $T_c$. The increase in $T_c$ above the value of bulk niobium (9.25 K) is only observed once the tetragonal phase exhibits fully resolved splitting of the (200)/(004) reflections, suggesting that a minimum degree of crystallographic ordering is needed, in agreement with EBSD experiments.

The absence of the cubic δ-NbN phase in our samples is indicative of a deficiency of nitrogen at the surface. By comparing with previous results of nitrided Nb prepared in a conventional furnace [41], an increase in the nitrogen pressure above 2.50 bar (our present set-up limit) during laser processing would further improve the superconducting properties of the niobium nitride layers generated by irradiation of bulk Nb using IR ns-lasers.

**Acknowledgments**

This publication is part of the projects PID2023-146041OB-C21 (funded by MICIU/AEI/10.13039/501100011033 and ERDF/EU) and T54-23R (funded by Gobierno de Aragón). J. Frechilla acknowledges support from Gobierno de Aragón through predoctoral contracts. The authors would like to acknowledge the use of Servicio General de Apoyo a la Investigación-SAI (Universidad de Zaragoza) and the Spanish National Facility ELECMI ICTS, node "Laboratorio de Microscopías Avanzadas (LMA)" at "Universidad de Zaragoza".

**Data availability**

The data that support the findings of this article are openly available at https://doi.org/10.5281/zenodo.19254032

# Supporting information:

# Laser nitriding of bulk niobium: phase evolution and superconducting behaviour

J. Frechilla[1], A. Frechilla[1], G.F. de la Fuente[1], A. Larrea[1], L.A. Angurel[1], E. Martínez[1]

1. Instituto de Nanociencia y Materiales de Aragón, INMA, CSIC-Universidad de Zaragoza, María de Luna, 3, 50018 Zaragoza, Spain

## S1. XRD patterns

In this supplementary section, the X-ray diffraction (XRD) patterns of the same samples presented in Figure 1 of the main text are shown over an extended angular range (30–80°). The samples are named as in the manuscript (Table 1). Specifically, the samples of Figure S1A (S1, S2, S3 and S4) were processed under different nitrogen pressure (0.15, 0.50, 1.50 and 2.50 bar, respectively) with fixed $F_{2D} \approx 7.5$ kJ/cm$^2$; and those in Figure S1B (S5, S6, S7 and S9), at 2.50 bar and $F_{2D} \approx$ 56, 113, 528, 1056, kJ/cm$^2$, respectively.

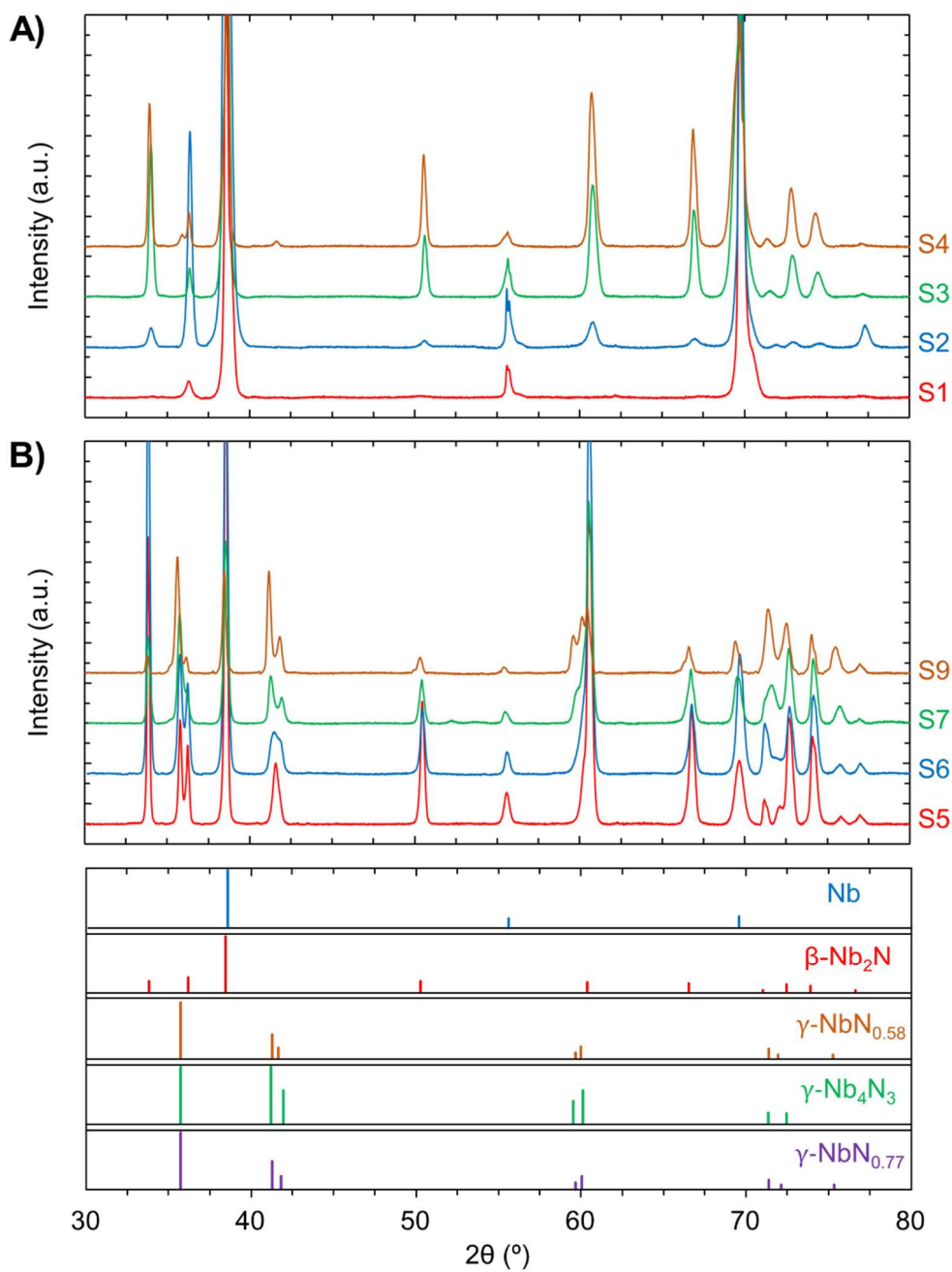

**Figure S1**. XRD extended patterns (range 30 – 80°) of selected laser-processed samples showing the evolution of the niobium nitride phases by modifying A) the nitrogen pressure for $F_{2D}$ = 7.5 kJ/cm$^2$; and B) the 2D accumulated fluence for $P_{N2}$ = 2.50 bar (see text). The peaks corresponding to relevant phases (see text and Table S1) are plotted below the experimental patterns.

As discussed previously, the evolution of the diffraction patterns confirms the formation of nitride phases as nitrogen pressure and 2D fluence increase. The Nb substrate and the hexagonal β-$Nb_2N$ phase can be clearly identified throughout the full angular range, with well-defined reflections. A more detailed analysis of the extended diffraction range reveals that the tetragonal phase does not perfectly match the γ-$Nb_4N_3$ phase. In particular, the measured peak at 2θ ≈ 75.4° is not reported for this phase, but it can be indexed as the (224) reflection of γ-$NbN_{0.58}$ or γ-$NbN_{0.77}$, for example (see table S1). This suggests that the tetragonal phase would correspond to a γ-$Nb_4N_{3\pm x}$ type structure with compositional variations across the nitride layer. Reflections corresponding to the cubic δ-NbN phase have not been observed.

Table S1: Relevant phases for XRD and EBSD analysis of the laser nitrided niobium samples

| Phase | Crystal System | a, b, c  [Å] | Space Group | Database #ID |
|---|---|---|---|---|
| Nb | Cubic | 3.30, 3.30, 3.30 | 229 | ICSD #76554 |
| $Nb_2N$ | Hexagonal | 3.06, 3.06, 4.96 | 174 | ICSD #31165 |
| $NbN_{0.58}$ | Tetragonal | 4.39, 4.39, 8.69 | 139 | ICSD #40075 |
| $Nb_4N_3$ | Tetragonal | 4.38, 4.38, 8.63 | 139 | ICSD #76389 |
| $NbN_{0.77}$ | Tetragonal | 4.39, 4.39, 8.66 | 121 | ICSD #66304 |

## S2. EBSD analysis

In this section we give additional details about the EBSD experiment and analysis. The maps were acquired at an accelerating voltage of 15 or 20 kV, with a pixel size of 0.15 ± 0.02 μm for the different samples. For the indexing process, the crystallographic phases Nb, β-$Nb_2N$ and γ-$NbN_{0.77}$, representing the tetragonal phase of table S1, were selected in the Aztec software. γ-$NbN_{0.77}$ was chosen because it provides the best fit to the tetragonal phase reflections observed in the XRD patterns (Section S1). We also substituted the γ-$NbN_{0.77}$ phase with other similar tetragonal phases, but we did not observe significant changes in the EBSD maps.

The number of bands used for the indexing was between 8 and 10 and approximately 40 reflectors were considered for the process for each phase. Calibration refinement was performed using β-$Nb_2N$ in all cases at an approximately central point in the map.

Excluding the epoxy area, the percentage of indexed pixels in the samples is ≈ 93 ± 2%, indicating good indexing quality and minimal presence of unindexed points, most of them located at the interfaces between different phases. MAD (Mean Angular Deviations) values were in the range of 0.55 - 0.80 for Nb, 0.55 - 0.65 for β-$Nb_2N$, and 0.60 - 0.75 for γ-$NbN_{0.77}$ (mean values, with std. dev. of about 0.15 - 0.20), confirming a good fit of the Kikuchi patterns for the selected phases.

As an example, Figure S2 shows three experimental Kikuchi patterns obtained in sample S7, corresponding to the three analysed phases: Nb, β-$Nb_2N$ and γ-$NbN_{0.77}$, together with the corresponding indexed solution.

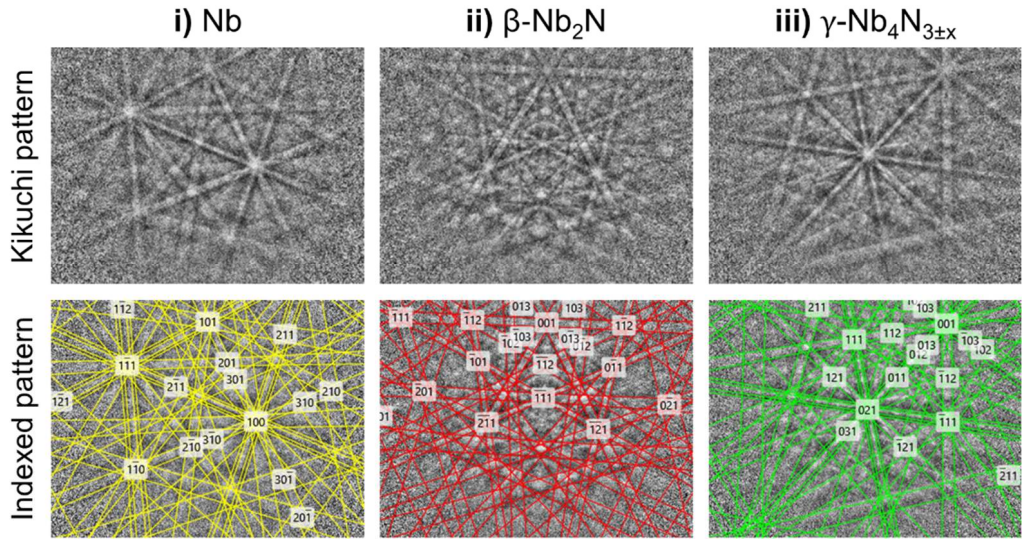

**Figure S2.** Kikuchi patterns of sample S7 ($F_{2D}$ = 528 kJ/cm$^2$) showing the experimental Kikuchi pattern and the corresponding indexed solution for i) Nb, ii) β-Nb$_2$N and iii) γ-Nb$_4$N$_{3\pm x}$.

The EBSD maps obtained for samples S6, S7 and S9 are shown in figures S3, S4 and S5, respectively. These maps are complementary to the results shown in Figures 4 and 5 of the main text. Here we have included the three IPF maps (for all-phases and only the Nb-phase), and the Kernel Average Misorientation (KAM) maps. For clarity, the images in these figures are overlaid on the FSE maps to improve visualization. For the KAM maps a 3×3 square filter and a threshold angle of 5° were used. For all studied samples, these maps reveal that the residual stress (green and yellow colours in the images) are concentrated mainly in the Nb matrix, near the interface with the nitride β-grains. These stresses are attributed to volumetric expansion during nitride formation combined with steep thermal gradients inherent to laser processing, leading to a highly deformed interfacial region.

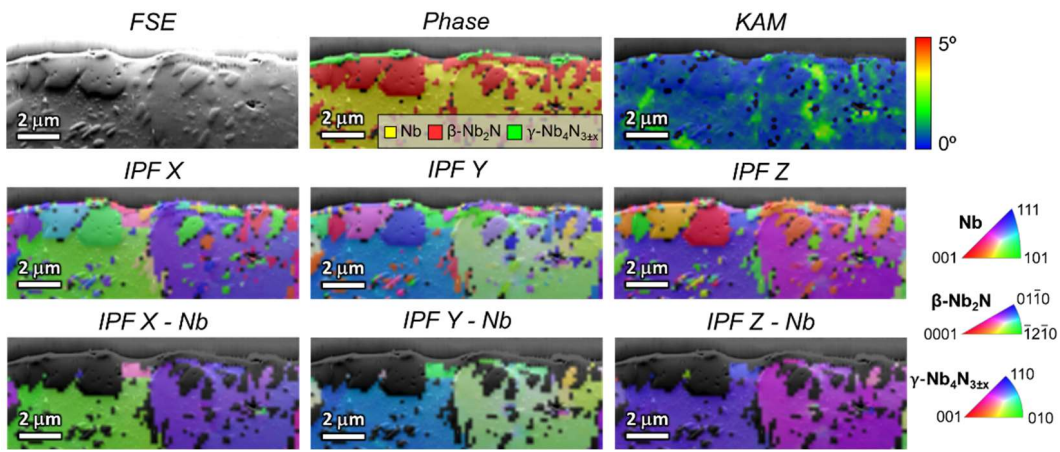

**Figure S3.** EBSD maps of sample S6 (2.50 bar, $F_{2D}$ = 113 kJ/cm$^2$) showing FSE, Phase, KAM, IPF X, IPF Y and IPF Z. The IPF X, IPF Y and IPF Z maps corresponding to the Nb-phase are also shown.

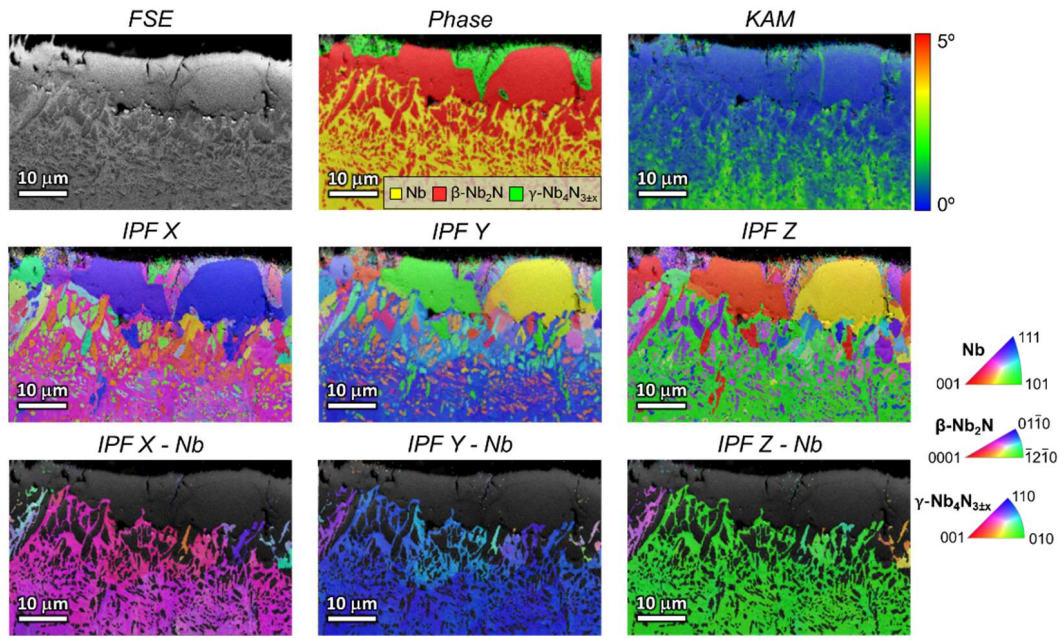

**Figure S4.** EBSD results of sample S7 (2.50 bar, $F_{2D}$ = 528 kJ/cm$^2$) showing FSE, Phase, KAM, IPF X, IPF Y and IPF Z maps. The IPF X, IPF Y and IPF Z maps corresponding to the Nb-phase are also shown.

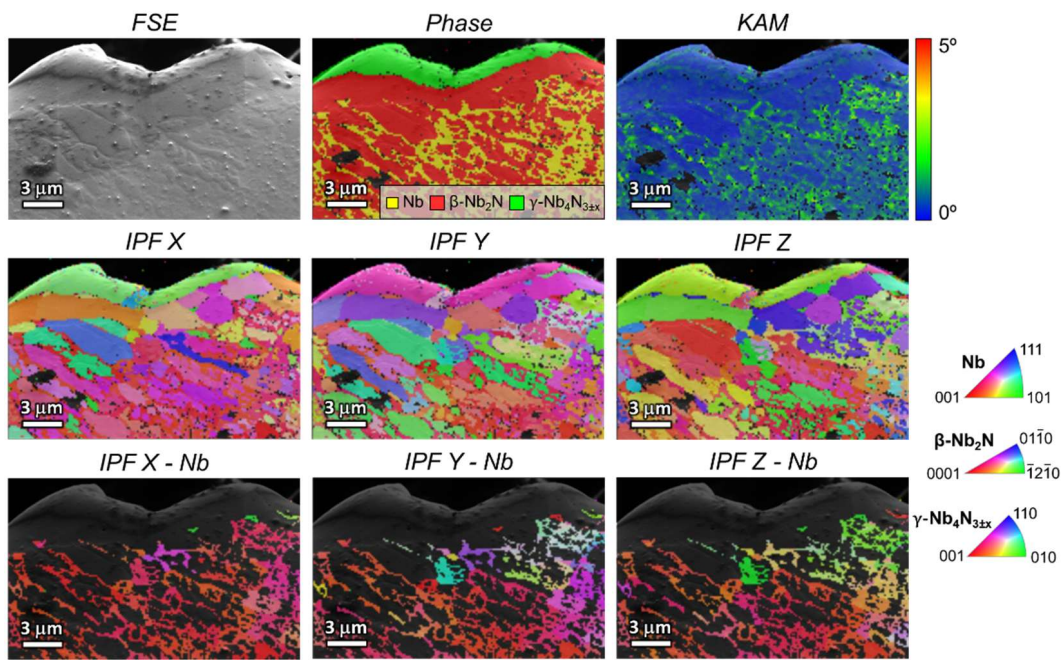

**Figure S5.** EBSD maps of sample S9 (2.50 bar, $F_{2D}$ = 1056 kJ/cm$^2$) showing FSE, Phase, KAM, IPF X, IPF Y and IPF Z. The IPF X, IPF Y and IPF Z maps corresponding to the Nb-phase are also shown.

26